# Principles of seed banks: complexity emerging from dormancy

Lennon JT[1], den Hollander F[2], Wilke-Berenguer M[3], Blath J[4]

[1] Indiana University, Department of Biology

[2] Universiteit Leiden, Mathematical Institute

[3] Humboldt-Universität zu Berlin, Institute of Mathematics

[4] Technische Universität Berlin, Institute of Mathematics

**Correspondence:** Jay T. Lennon, 1001 E. 3rd St. Bloomington, Indiana, 47401, USA,

lennonj@indiana.edu

# ABSTRACT

Across the tree of life, populations have evolved the capacity to contend with suboptimal conditions by engaging in dormancy, whereby individuals enter a reversible state of reduced metabolic activity. The resulting seed banks are complex, storing information and imparting memory that gives rise to multi-scale structures and networks spanning collections of cells to entire ecosystems. We outline the fundamental attributes and emergent phenomena associated with dormancy and seed banks, with the vision for a unifying and mathematically based framework that can address problems in the life sciences, ranging from global change to cancer biology.

# INTRODUCTION

Seed banks are reservoirs of metabolically inactive individuals. Collectively, the genetic, phenotypic, and functional diversity of these dormant individuals can influence the behavior of biological systems in important and wide-ranging ways. Almost all biological entities can generate a seed bank, but perhaps the most iconic and well-studied examples come from plants (Fig. 1B). Thousands of years ago, major transformations in culture and society arose when humans began to harness the diversity and longevity of plant seeds[1]. Much later, Darwin's curiosity was piqued by seed banks when more than 500 types of plants germinated from only three tablespoons of mud collected from the margins of an English pond. This led him to ponder the ways in which animals might promote the movement of dormant propagules, and how this in turn might affect the geographic distribution of species[2]. Since then, a rich body of theoretical, comparative, and experimental work has provided critical insight into the stabilizing role of seed banks for the diversity and eco-evolutionary dynamics of plant populations and communities[3-7]. Inspired by observations in natural systems, seed-banking vaults have been engineered around the world to help preserve biodiversity and ensure food security through the long-term storage of rare plant cultivars that are at risk of extinction owing to rising temperatures, disease outbreaks, and other natural disasters[8].

Yet, seed banks are by no means restricted to the world of plants. Dormancy, which is essential for seed-bank formation, has independently evolved numerous times throughout Earth's history and is even considered in some origins-of-life models[9,10]. In today's inventory of biodiversity, dormancy is integrated into diverse modes of survival and reproduction across domains of life (Fig. 1). For example, dormancy is achieved via latency in viruses[11], sporulation by bacteria and fungi[12], cyst formation among protists[13], estivation by amphibians[14], dauer stages by worms[15], cryptobiosis in rotifers[16], diapause by insects[17], torpor in birds[18], quiescence in fish[19], delayed blastocyst formation among marsupials[20], and hibernation in mammals[21]. The principles of dormancy and seed banks have even been extended to stem-cell dynamics[22], tumor persistence[23], wound healing[24], immune functioning[25], and neural activity of brains[26]. Despite their prevalence across seemingly disparate systems and scales, seed banks are governed by a shared set of dormancy-related attributes and processes that allow individuals to disperse in time (Fig. 2). As an example of convergent evolution, seed banking represents a common solution to some of the major challenges faced by nearly all forms of life.

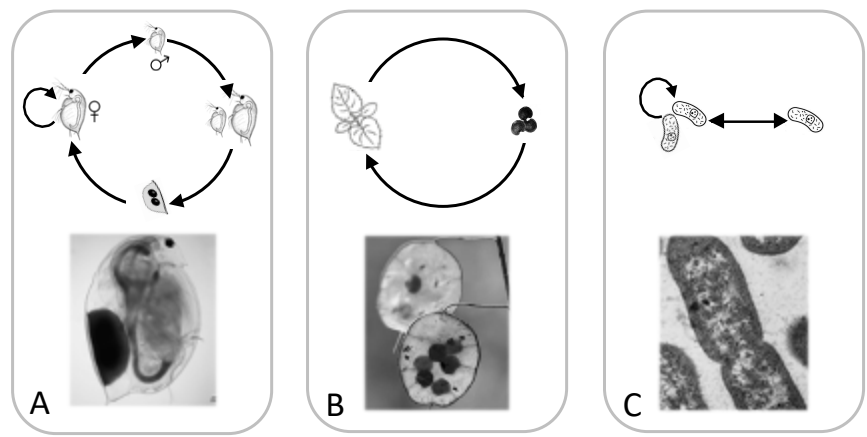

**Fig. 1.** Seed banks develop among diverse taxa with different life histories and reproductive modes. **A**. *Daphnia* are planktonic crustaceans with parthenogenic reproduction, where females generate offspring without fertilization. When stressed, some species produce males, who in turn fertilize females, leading to the production of dormant resting stages (ephippia) that can persist for extended periods of time in aquatic sediments. **B**. Annual plants produce seeds at the end of a growing season, which are deposited into the soil. As a well-recognized example of bet hedging, some fraction of these seeds delay germination despite optimal conditions. **C**. Among many groups of microorganisms, individuals can enter and exit from a dormant state independent of reproduction and without the need of generating physical resting structures.

Seed banks are important buffers that impart resiliency on biological systems. By entering a dormant state, individuals enjoy protection from inhospitable conditions that would otherwise reduce a population's fitness. When lucky, at least some of these individuals will awake in the future when the environment is more conducive for growth and reproduction. Given its consequences for evolution and ecology, considerable effort has been devoted towards characterizing the mechanisms that control seed-bank dynamics. For some populations, the success of dormancy is achieved by responsive transitioning between metabolic states where the finely tuned regulation of signals, hormones, and even neural circuits[27] allows organisms to interpret, integrate, and respond to information about their internal and external environment in a deterministic fashion. For other populations, the ability to enter dormancy and delay reproduction, despite developmental preparedness, involves stochastic bet hedging, which is effective in reducing the probability of extinction and maximizing fitness in unpredictable environments[28]. While the underlying details can be critical for an in-depth understanding of how dormancy operates within a particular organism or population, it is not our objective to provide a comprehensive description of the molecular and physiological factors underlying dormancy among broad lineages of plants, animals, and microorganisms. Rather, our goal is to gain new insight into seed banks by identifying common features and critical variables that are amenable to systems theory and mathematical modeling. Approaches that embrace stochastic processes and nonlinear interactions in large systems are particularly well poised for understanding the emergence of complex seed-bank patterns across scales. Our hope is that such a framework will not only lead to generalizable and predictive knowledge, but also foster cross-disciplinary efforts by alleviating confusion stemming from the lexicon that is commonly associated with dormancy and seed banking[29].

It has been more than a half-century since theoreticians took the initial steps to formalize a seed-bank theory[30,31]. Motivated by observations in nature, macroscale phenomena have often been modeled as dynamical systems, providing insight into the adaptiveness of seed banks and optimality of dormancy in fluctuating environments[4,32-34]. In recent years, this foundational body of research has expanded into other disciplines, inspiring novel mathematical and computational approaches that combine individuality, stochasticity, and memory, while allowing for descriptions of probabilistic and deterministic scaling limits[35-37]. Taken together, these complementary perspectives present an opportunity for unification and universality, while also raising intrinsic and challenging problems for mathematicians. As such, the objective of our paper is to introduce and define the core seed-bank attributes and dormancy-related processes that can be studied and modeled across systems and scales. We explore the consequences of seed banks for key biological phenomena, including evolutionary processes, population dynamics, species interactions, and ecosystem functioning. Along the way, we rely on Boxes to highlight recent mathematical developments pertaining to dormancy and seed banks with an emphasis on stochastic individual-based models and inferences that can be gained from coalescence theory. We conclude with a discussion on the frontiers and challenges of the seed-bank framework for practical issues in the life sciences, including aspects of human health and global-change biology.

## FUNDAMENTALS OF SEED BANK THEORY

In this section, we identify the core attributes and processes that are essential for establishing a general framework for seed banks and their dynamics. We emphasize that seed banks have multiscale properties. For example, metabolic transitions occur on time scales spanning seconds to millennia, and operate across levels of biological organization, ranging from genes within individuals to energy flow within food webs. While some seed-bank properties are tractable to empirical measurement, others can be more challenging to describe, which creates opportunities for investigations that leverage theory and modeling.

Seed-bank attributes: The primary object of a seed bank is the pool of dormant individuals. A critical attribute of this pool is its size, which is simply the sum of all viable but inactive individuals or biomass (Fig. 2). In some cases, seed banks are large, making up nearly all of the individuals in a system. For example, 90 % or more of all the bacteria and fungi living in soil can be dormant[12]. In marine sediments alone, there are an estimated $10^{29}$ endospores, suggesting that microbial seed banks are significant at the global scale[38]. In many plant systems, dormant individuals attain densities that exceed the abundance of actively growing individuals by orders of magnitude[39].

In addition to their absolute and relative sizes, seed banks have other important attributes. For example, dormant individuals can be assigned to different "classes", which might represent genotypes in a population or species in a community (Fig. 2). There are many ways in which the diversity of these classes can be quantified[40]. By applying commonly used metrics from ecology and information science, one can estimate the $\alpha$-diversity of the seed bank as the total number of different classes (i.e., richness), the equitability in abundance of the different classes (i.e., evenness), or the degree of entropy among classes within a sample (i.e., Shannon Index). Similarly, the concept of $\beta$-diversity can be applied to seed banks. While often used to characterize variation among distinct samples in time or space, the compositional similarity between classes in the active and inactive pools of a sample can also be quantified. For example, in the study of an alpine grassland there was only 20 % overlap in the species found in the aboveground plant community compared to the belowground seed bank[41]. As we will discuss in the following sections, the ability to characterize the metabolic activity of individuals has important implications for understanding dynamics and emergent phenomena in a range of biological systems.

Dormant individuals possess other important characteristics that affect seed-bank dynamics. For example, they may have unique evolutionary histories, demographic properties, or functional traits that affect aspects of organismal performance, such as maximum growth rate, grazing tolerance, drug resistance, mutation rate, and light requirements, etc. When combined with pool sizes and transition probabilities, these characteristics can influence seed-bank turnover and other emergent phenomena.

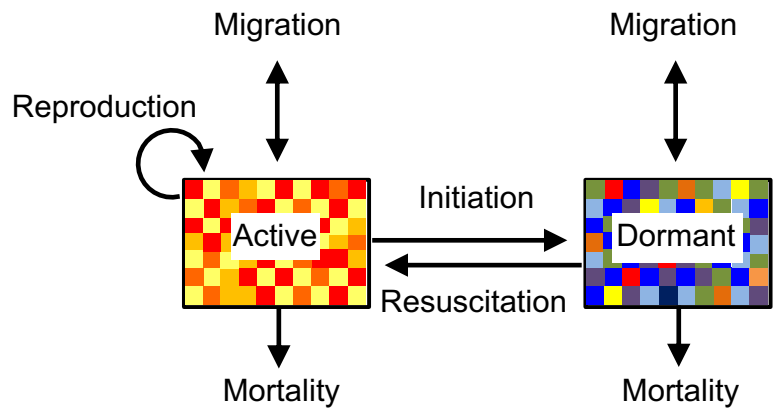

**Fig. 2.** Primary attributes (boxes) and transitions (arrows) in a generalized seed-bank model. In this example, the sizes of the active and dormant pools are made up of an equal number of individuals ($N$ = 70) belonging to different classes (colored squares), which may represent genotypes within a population or species within a community. In the active pool, individuals can be gained through reproduction and lost through mortality. In the dormant pool, there is no reproduction and mortality of inactive individuals is typically assumed to be much lower than for active individuals, which is reasonable for many but not all taxa that invest in long-lived seed banks. In addition, pool sizes are influenced by stochastic or deterministic transitions between metabolic states (i.e., initiation and resuscitation), which determine the size and rate at which pools undergo turnover. In terms of $\alpha$-

diversity, the richness of classes in the dormant pool ($S_d$ = 9) is greater than the richness in the active pool ($S_a$ = 4). In terms of β-diversity, the active and inactive pools are 82 % dissimilar based on the abundance-weighted Bray-Curtis metric: $\sum_{k=1}^{S}[x_{ak} - x_{dk}]/\sum_{k=1}^{S}(x_{ak} + x_{dk})$, where $x_{ak}$ and $x_{dk}$ correspond to the abundance of class $k$ in the active ($a$) and dormant ($d$) pools, respectively, and $S$ is the number of classes contained in the pools. Seed-bank attributes can also be influenced by migration, especially when dormancy facilitates the dispersal and colonization of individuals in a regional landscape.

Seed-bank transitions: Ultimately, the primary attributes of a seed bank are governed by the transitions of individuals between metabolic states. Therefore, it is crucial to consider the different ways that individuals move in and out of dormancy. In unpredictable environments, bet hedging strategies evolve[28]. These strategies require giving up potential gains that could be attained in one type of environment to avoid risks associated with another type of environment. For example, in what is referred to as conservative bet hedging, a plant might "always play it safe" by producing a smaller number of high-quality seeds, regardless of environmental conditions[28]. This generalist tactic increases the multiplicative growth rate (i.e., geometric mean fitness) of the population by decreasing the variance in performance among individuals over time[42]. An alternate strategy is diversified bet hedging, which metaphorically is interpreted as "not putting all of your eggs in one basket". In this case, a plant might spread risk by generating seeds with different germination phenotypes, which increases geometric mean fitness, not by reducing variance among individuals, but by reducing the correlation in performance among offspring[43,44]. Finally, a stochastic variation of diversified bet hedging is adaptive coin-flipping. As before, a plant produces seeds with different germination phenotypes, but the exact reproductive strategy is chosen randomly and independently by each individual in subsequent generations[43,45]. Outside of plants, stochastic bet hedging has attracted attention in microbial systems owing to its role in generating phenotypic heterogeneity[46-49]. For example, persister cells are thought to confer tolerance to stressors like antibiotics, not through the acquisition of mutations, but through stochastic gene expression that causes a small fraction of the population to enter a dormant state[50]. Similarly, the scout hypothesis considers stochastic resuscitation from dormancy in a large population where, independent of any known cues, individuals randomly wake up and sample their environment[51,52]. If conditions are bad, then the individual may die or go back to sleep. If conditions are good, then the individual can communicate and coordinate resuscitation with other members of the population in a density-dependent manner.

While nature is often uncertain, there are features of an organism's environment that change relatively slowly or in a fairly predictable manner (e.g., photoperiod, temperature, rainfall). Under such conditions, instead of relying on bet hedging, many populations have evolved responsive mechanisms that regulate how and when individuals enter and exit dormancy. Although this form of phenotypic plasticity can diminish the effect of noise on decision making, it is still a costly strategy. For example, a sensing system requires that individuals invest limited resources and time towards the maintenance of network components, which include receptors, readout molecules, and energy to perform work[53]. These allocation burdens, combined with environmental fluctuations, can set up trade-offs that have implications for population performance. But at least in some environments, the investment into responsive transitioning means that organisms are better equipped to make decisions based on internal and external cues that can improve fitness outcomes and even give rise to anticipatory behaviors. We discuss quantitative approaches to evaluating the optimality of dormancy transitioning with respect to environmental switching in Box 1.

Often, it is justifiable to model seed banks using a two-compartment model where individuals make one-step transitions between endpoint metabolic states (Fig. 2). However, dormancy can be more complicated. For example, in some plant systems seeds move through a series of metabolic states en route to germination[54]. Progression through these stages can be determined by endogenous factors such as the levels of phytohormones (e.g., gibberellin and abscisic acid) in seeds[55], but also by exogenous requirements such as physical abrasion of the seed coat, or even passage of the dormant propagule through the mammalian gastrointestinal tract[56]. Thus, at any given point in time an individual sampled from the seed bank may rest in a shallow or deep state of dormancy[39]. Even for some single-celled bacteria, dormancy requires transit through a sequence of developmental stages. In populations of *Bacillus* precise regulation of multiple transcription factors leads to an irreversible commitment to dormancy[57], which has important implications for fitness in time-varying environments. On average, it takes a *Bacillus* cell eight hours to create a functional spore [58]. During that time, the same individual with a 20 min doubling time could give rise to $10^7$ offspring provided that conditions would be optimal for growth. Meanwhile, other types of microorganisms fall into dormancy without needing to generate a distinct resting stage[12]. In such cases, it might be more appropriate to consider individuals lying along a continuum of metabolic activity that reflects quantitative variation in physiological characteristics such as resource quota[59], ribosome number[60], or the degree of DNA methylation[61]. In other words, special features and transitions may be needed to accurately model seed-bank dynamics in some systems.

**Box 1. Optimal dormancy strategies in fluctuating environments**

Understanding the costs and benefits of dormancy is important for the ecological and evolutionary modeling of seed banks. Central to this aim are questions pertaining to the optimality of deterministic and stochastic strategies in relation to the dynamics of a population's environment. To illustrate, consider an idealized population reproducing in discrete generations according to a two-type branching process. One type corresponds to the active state of an individual and the other type corresponds to its dormant state. Further, assume that the process evolves in a randomly fluctuating environment with a good condition that is favorable for reproduction and a bad condition that is unfavorable for reproduction. The per generation switching probabilities between environments are denoted by $s_1$ (good → bad) and $s_2$ (bad → good). First, we can consider stochastic transitioning, where individuals move between metabolic states irrespective of the environmental condition (good vs. bad). Depending on the specifics of the model, this form of stochastic transitioning can be interpreted as diversified bet-hedging or adaptive coin flipping[28]. Second, we can consider responsive transitioning, where individuals enter a dormant state when conditions are bad and resuscitate into the active state when conditions are good. Last, we can consider anticipatory (or prescient) transitioning, where individuals enter a dormant state during good conditions in preparation of impending bad conditions. The optimality of these dormancy strategies can be explicitly evaluated in relation to environmental conditions and underlying parameters[62]. Here, fitness is measured in terms of the maximal Lyapunov exponent of the populations, which can be interpreted as long-term geometric mean fitness[63,64]. As shown in the figure below, responsive transitioning has positive fitness and is optimal among all strategies when the environment slowly changes (red-colored regions). Stochastic transitioning is optimal and has positive fitness when there are moderate fluctuations in environmental conditions (blue-colored regions). Anticipatory transitioning is most adaptive in environments with rapid and almost predicable changes between good and bad conditions that occur on the time scale of a generation (green-colored regions). As expected, dormancy is not favorable when environmental conditions are good most of the time (yellow-colored region). This qualitative picture holds for a range of parameters, and the model can be extended to incorporate reproductive trade-offs, resource limitation, and different investment costs. For example, when dormancy becomes too costly, the spontaneous transitioning strategy becomes less optimal (shrinking of the blue-colored regions).

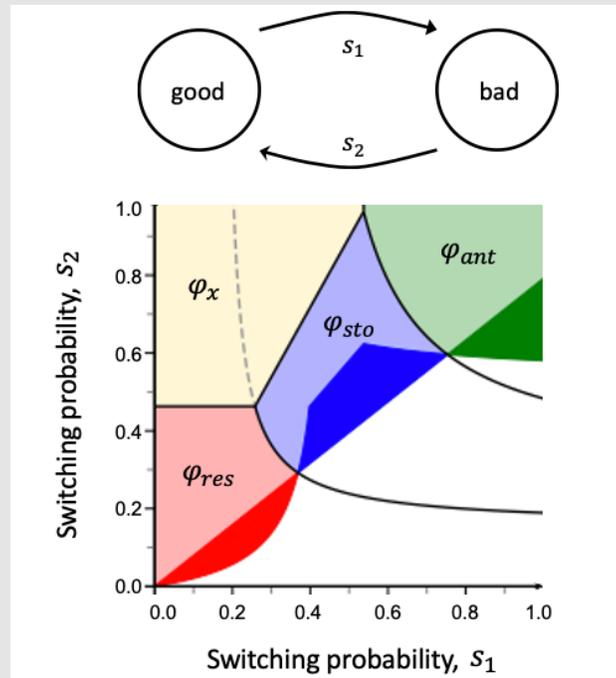

**Upper panel:** Switching dynamics in a randomly fluctuating environment. When the values of $s_1$ and $s_2$ are small, the environment is relatively stable. When the values of $s_1$ and $s_2$ are large, the environment rapidly fluctuates between good and bad environmental conditions, changing almost every generation. **Lower panel:** optimality regions corresponding to different transition strategies[62]. Specifically, $\varphi_{sto}$, $\varphi_{res}$, and $\varphi_{ant}$ indicate regions where a population employing stochastic (blue), responsive (red), and anticipatory (green) strategies are optimal and have positive fitness, while $\varphi_x$ depicts the region where a population that does not engage in dormant has the highest (and positive) fitness (yellow). Within a color family, the dark-colored regions indicate that the optimal strategy is the only one with positive fitness. In the light-colored regions, there is at least one other strategy with positive, but strictly lower than optimal, fitness. In the white regions, where environmental conditions are
13

stably bad (large $s_1$ and small $s_2$), all strategies, including the optimal one, have negative fitness. Precise model definitions and parameters can be found elsewhere[62].

Physical features of seed banks: The physical manifestation of seed banks has important implications for understanding dormancy dynamics and emergent phenomena. Compared to members of the active pool, dormant individuals often take on different sizes, densities, and motilities. These morphological distinctions can facilitate the transport of propagules to new environments, especially when dormant individuals act as agents of dispersal for future generations, as in plants and zooplankton (Fig. 1A, B). Although it is well documented that most offspring will develop within close vicinity to their natal site, dormancy has the potential to expand the dispersal kernel and the geographic range of some species[65].

In many systems, however, active and dormant individuals have similar morphological features and occupy the same habitat. For example, in the surface waters of lakes and oceans, phytoplankton are at the mercy of currents, which can lead to the homogenization of individuals within water masses, irrespective of their metabolic status[66]. In such environments, if active and dormant individuals cannot readily be distinguished, then eco-evolutionary patterns may be obscured or otherwise go undetected, which potentially leads to biased interpretation of empirical observations[36,67].

Yet in other systems, seed banks play an important role in the development and construction of biophysical structures. For example, dormant individuals are integrated into the organization of multicellular biofilms and tumors. In such systems, the mixture of active and dormant cells create structures that alter fluid dynamics, oxygen availability, and the generation of heat, which in turn can influence toxin production, communication networks, and even the lateral transfer of DNA[68,69]. In this way, variation in the metabolic activity of neighboring individuals can influence the spatial configuration of seed banks and generate feedback that influences system behavior.

## SEED BANKS AND EMERGENT PHENOMENA

Once established, seed banks can lead to the emergence of interesting patterns and dynamics that do not necessarily align with expectations of classical theory. This is because seed banks contain structure, which imparts memory on the system. Through the process of delayed resuscitation, individuals from the past can arise in the future and influence behavior, including equilibria and stability, but also multiscale organization and feedback. In this section, we outline some of the fundamental processes that are modified by seed banks and discuss how dormancy influences major evolutionary and ecological patterns.

Evolution with a seed bank: Seed banks modify the fundamental forces of evolution acting upon a population. For example, dormancy can influence genetic diversity by altering the input of mutations arising from point mutations, frameshifts, and duplication events. On the one hand, we might expect the number of mutations to be relatively high in dividing cells, since errors are often created during genome replication[70]. On the other hand, many mutations generated by dividing cells can be efficiently repaired by mechanisms including polymerase-based proofreading during DNA replication and mismatch repair immediately following DNA synthesis[70]. These error-correcting mechanisms are less efficient in non-dividing cells, perhaps making them more susceptible to DNA damage caused by background radiation, free oxygen radicals, and starvation, which can result in elevated mutation rates[71]. While non-trivial to quantify, mutation rates per unit time appear to be about 10- to 20-fold higher for dividing cells compared to non-dividing cells[72,73]. In addition, dormancy may affect the types of mutations entering a population. For example, the mutational spectrum of actively growing microbes tends to be enriched with insertions, as well as transitions and transversions that result in AT bias[74]. In contrast, quiescent yeast populations were found to exhibit less AT bias and a higher occurrence of deletions[73]. Together, these findings suggest that dormancy should have strong effects on the supply of mutations, with important implications for population genetics and molecular evolution.

Once generated, the fate of a mutation is affected by the seed bank. Under neutral conditions, genetic drift can lead to the fixation of alleles through random sampling. The resulting loss of genetic diversity is especially important in small populations. Because seed banks increase the effective population size $N_e$, it should take longer for alleles to reach fixation via genetic drift[75]. Such effects are reflected in the site frequency spectrum (SFS), a relationship that describes how segregating alleles are distributed among loci in a population. The shape of the SFS is often interpreted with respect to demographic processes, such as bottlenecking, population expansion, and immigration, which can influence the frequency of rare vs. common mutations in a population. However, the SFS is also affected by seed-bank processes [35,76-78]. For example, coalescent theory predicts that the SFS will contain more singletons and segregating sites, but the degree to which seed banks retain genetic diversity depends on the average amount of time that individuals spend in a dormant state (Box 2). Such findings not only have important implications for quantifying genetic diversity, but also for properly interpreting the processes that give rise to these estimates.

Seed banks alter the strength of natural selection, which has important consequences for rates of adaptation. Like genetic drift, natural selection is a process that reduces genetic diversity by either purging deleterious mutations or allowing individuals with beneficial mutations to sweep through a population. In the presence of a seed bank, mildly deleterious alleles may be retained in a population because they are not seen by natural selection. Such logic leads to the prediction that seed banks may slow down rates of evolution[5], which has been supported by field observations[79]. However, when dormant variants resuscitate, perhaps under different environmental conditions, they contribute to the standing genetic variation of a population. In this way, seed banks may allow populations to capture reproductive opportunities, which should accelerate rates of adaptive evolution[37,75]. In a dynamic environment, selection with a seed bank may preserve genetic and phenotypic diversity. For example, the maintenance of a color polymorphism in an annual plant population was explained by a diffusion approximation model that accounted for fluctuating selection involving the storage and resuscitation of individuals from the seed bank[80]. By considering population-genetic processes, like selection, at larger spatial scales, it has been hypothesized that seed banks provide an anchoring effect that buffers populations against gene flow in the landscape[81].

Ultimately, seed banks have the potential to shape macroevolutionary phenomena. Coalescent models predict that long-term dormancy can alter genealogical patterns and our ability to estimate the time to the most recent common ancestor (Box 2). In nature, such patterns may be influenced by the seed bank's ability to reduce the frequency of extinction events. But, persisting in a non-replicating state for prolonged periods of time may also alter rates of lineage diversification, even though evidence to support this notion is mixed. In a comparative analysis of bacterial genomes, amino acid substitution rates for endospore-forming vs. non-endospore-forming strains were comparable, suggesting that dormancy may not leave a strong signature on rates of molecular evolution[82]. Yet, in a much larger analysis, lineages of endospore-forming bacteria diverged more slowly than lineages that had lost the capacity to form long-lived resting stages[83]. In a study of more than 14,000 plant taxa, the effects of seed banks on lineage diversification appear to be sensitive to the type of dormancy[84]. For example, plants engaging in physical dormancy had very low (negative) rates of net diversification owing to high extinction rates. In contrast, plants that rely on physiological or morphological dormancy had higher rates of speciation, which led to elevated (positive) rates of net diversification compared to species that lacked dormancy altogether. Together, these findings suggest that the influence of seed banks on macroevolutionary processes may vary among taxa, perhaps reflecting differences in the underlying mechanisms of dormancy.

**Box 2. Coalescent theory for seed banks**

Coalescents are stochastic processes that are widely used to model genealogies on evolutionary time scales. They are indispensable for understanding the interplay of genetic drift with other evolutionary forces, including mutation, selection, recombination[85]. Below we describe classic and more recently developed coalescent models in the context of seed-bank processes. While some models already allow for evolutionary inference[76,77,86], others form the basis for future development of testable predictions.

1. The Kingman coalescent: Introduced nearly 40 years ago, "the coalescent" has played an important role in advancing population-genetic theory and application[87]. For describing the genealogy of a population under genetic drift, the coalescent is the standard null model for neutral evolution, since it arises as the universal scaling limit of the ancestral process in Wright-Fisher and Moran models, among others. Looking back in time, the coalescent approach makes pairwise mergers with ancestors at an exponential rate of 1 on the evolutionary time scale (panel A). Based on assumptions of a haploid Wright-Fisher model with a finite effective population size $N$, the coalescent generates a random genealogical tree, where time $t$ corresponds to $N_t$ generations in the finite population. This tree can be used to infer the expected time until the most recent common ancestor, which with a sample of size $n$ is always bounded by 2, corresponding to $2N$ generations in the underlying Wright-Fisher model. If mutations happen with probability $\mu$ per generation, then they will be visible on the coalescent with rate $\theta = 2N\mu$, providing the expected number of segregating sites in two lineages from the sample $n$. This can readily be estimated from DNA samples by using metrics like the Watterson estimator ($\hat{\theta}_w$). Since its inception, the coalescent has been used and modified by evolutionary biologists and mathematicians to explore a range of topics and processes. As we will see later, features of the coalescent (e.g., $\theta$) also play an important role in ecological models used to predict patterns of biodiversity, including the species abundance distribution (SAD) (Box 4).

2. <u>The stretched coalescent:</u> When individuals enter dormancy for a relatively short period of time, they give rise to what is called a weak seed bank. Because this increases the effective population size $N_e$, the resulting genealogy becomes "stretched"[86,88], such that pairwise mergers of ancestors occur less frequently. Specifically, if an individual spends on average $\beta$ generations in the seed bank, then the time to coalescence for two lineages is increased by a factor of $\beta^2$. The upper bound for the expected time to the most recent common ancestor therefore increases to $2\beta^2$, which again translates to $2\beta^2 N$ generations. Likewise, estimates of the coalescent mutation rate $\theta$ will be affected by a factor of $\beta^2$. However, other population genetic quantities, like the normalized site-frequency spectrum (nSFS) and the topology of the genealogy, remain unchanged (panel B).

3. The seed bank coalescent: When the time spent in a dormant state is longer, on the order of the effective population size, a strong seed bank emerges. This necessitates a distinction between active and dormant ancestral lineages in the genealogy. Only lineages of active individuals may coalesce, while mergers between dormant lineages are prohibited. As a result, the expected time to the most recent common ancestor increases further. Unlike in the stretched coalescent, where the time to the most recent common ancestor was extended by a constant, the time to the most recent common ancestor in the strong seed bank regime is unbounded in the sample size. The mechanisms underlying metabolic transitions can further modify coalescent statistics in non-trivial ways[35,89]. For example, when lineages transition between states independently of other events, the seed bank coalescent emerges (panel C). On the other hand, responsive or anticipatory transitions (Box 1) may cause lineages to enter and exit dormant states simultaneously, leading to a more generalized on/off coalescent pattern (panel D).

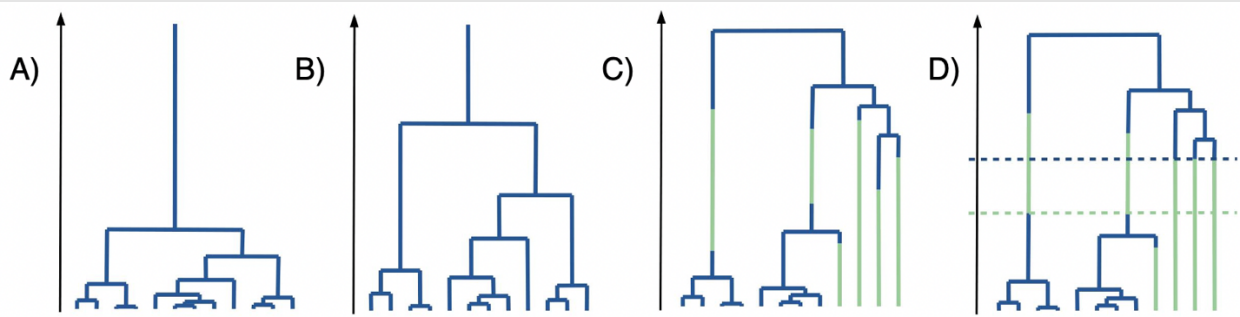

Genealogies with different seed bank attributes. **Panel A:** typical genealogy based on the Kingman coalescent without a seed bank. **Panel B:** genealogy from the stretched coalescent which arises when there is a weak seed bank effect. This leads to an increased amount of time for pairs of lineages to merge. **Panel C:** genealogy under a strong seed bank where lineages stochastically transition between active and dormant states, which leads to the seed bank coalescent. **Panel D:** genealogy under a strong seed bank where lineages undergo simultaneous transitioning, which captures behaviors where individuals respond to environmental cues, resulting in a more general on/off coalescent. Blue lines depict active lineages and green lines represent dormant lineages. Horizontal dotted lines in panel D reflect times of simultaneous transitioning.

Population dynamics with a seed bank: Seed banks have important implications for the dynamics and long-term stability of populations made up of individuals belonging to the same species. The combined use of models and empirical data has demonstrated that population dynamics are better predicted when dormancy is included, even if stages and transitions are sometimes difficult to quantify in nature[90]. For example, by using stochastic projection matrix models, it has been shown that species with longer life spans are less sensitive to variability in vital rates caused by fluctuations in environmental conditions[91]. Recent studies have begun to leverage the power of other modeling approaches, like the coalescent (Box 2), along with whole-genome data to infer the importance of dormancy on the demographic history of populations[77]. Together, these efforts have led to the prevailing view that, by temporarily suspending metabolic activities and distributing reproductive output over longer periods of time, seed banks reduce the probability that individuals will succumb to unfavorable conditions. As a consequence, important demographic features of a population, including its size and age-structure, are influenced by seed-bank dynamics[90,92].

The population-level consequences of seed banks are sensitive to the amount of time that an individual spends in a metabolically inactive state. While dormancy helps an individual conserve energy, there are still other basal metabolic demands that must be met, including costs associated with cellular homeostasis and the repair of macromolecular damage, which can contribute to diminished fitness associated with aging[93]. Other factors affecting the residence time of a seed bank include the compounded risk of consumption by predators, infection by pathogens, and physical burial[39,94]. Because of these factors, some seed banks turnover more rapidly than others. For example, less than 10 % of seeds belonging to an annual plant species in the Sonoran Desert persisted beyond five years[92], while radiometric dating revealed that diapausing zooplankton can hatch from lake sediments after more than 100 years[95]. Astonishingly, viable microorganisms can be recovered from ancient materials that are 100 million years old[96].

While dormancy has the potential to extend longevity, resuscitation must happen under natural conditions with sufficient frequency to influence the fitness and dynamics of a population. To this end, novel genome-barcoding approaches in combination with Bayesian methods have been used to characterize the variance of reproductive output associated with the resuscitation of dormant bacteria. Such approaches revealed that stochastic resuscitation drawn from a heavy-tailed distribution of variants resulted in a few individuals effectively winning a demographic sweepstake[97]. Not only can these long-range jumps diminish genetic diversity, they can also alter the underlying coalescent structure of a population[98,99]. Although genotypes from the past are not always well adapted to conditions of the future[100], resuscitation tends to benefit populations in many scenarios. In addition to minimizing competition among kin at similar developmental stages, recruitment of dormant propagules from the seed bank can buffer populations from various environmental stressors[95] and can be used to document coevolutionary Red-Queen dynamics between hosts and their parasites[101].

**Box 3. Emergence of dormancy under competitive pressure**

Dormancy is often investigated in the context of fluctuations that are driven by external forces. However, theory predicts that dormancy can also emerge in scenarios where fluctuations are internally generated[102-104]. For example, if competition is strong enough, then individuals can escape mortality associated with overcrowding by transitioning into a dormant state, conferring a benefit that offsets the reproductive costs associated with dormancy. Let us consider a stochastic individual-based model with two species. Species 1 is at equilibrium and reproduces at rate $\lambda_1 > 0$, which means that each individual divides after an exponential time independently of others. Individuals of the newly arriving Species 2 have a reduced rate of reproduction ($\lambda_2 < \lambda_1$) that reflects the cost of them being able to engage in dormancy. Further, assume that active individuals belonging to both species have a death rate that is less than $\lambda_2$. Then competition is proportional to the sum of all active individuals ($n_a$) belonging to Species 1 ($n_1$) and Species 2 ($n_{2a}$). This results in individuals from Species 1 being removed from the system due to competition at rate $n_a$. Because they can escape with probability $p$ into the dormant state, individuals from Species 2 are less sensitive to competition. Specifically, those individuals are removed from the community due to competition at rate $(1-p)n_a$ and transition into competition-induced dormancy at rate $pn_a$. Last, dormant individuals from Species 2 can resuscitate at rate 1. The invasion and subsequent fixation of dormancy-capable Species 2 is possible if and only if $\lambda_1 - \mu < (\lambda_2 - \mu)/(1-p)$[103]. While $\lambda_1 - \mu$ is always larger than $\lambda_2 - \mu$, due to the reproductive trade-off, this is balanced by the pre-factor $1/(1-p)$, which is always larger than one. In fact, for any choice of $\lambda_1$ and $\lambda_2$, it is possible to find an escape probability $p$ such that the dormant type has a chance to invade and lead to the competitive exclusion of Species 1. Thus, dormancy reduces the effects of competition, which compensates for lower reproductive rates, and has important implications for biodiversity and coexistence theory.

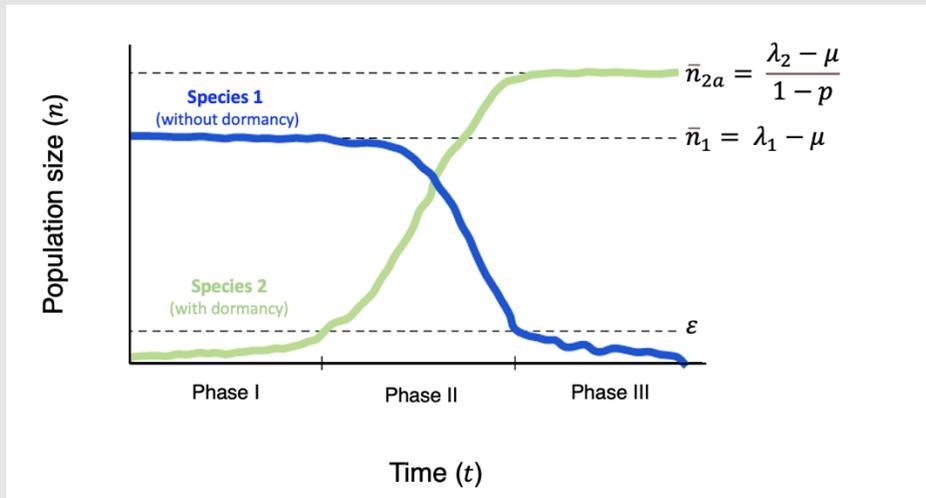

Invasion dynamics and fixation of a dormancy trait in a non-fluctuating environment despite reproductive trade off. In Phase I, individuals belonging to Species 1 are close to their equilibrium abundance ($\bar{n}_1 = \lambda_1 - \mu$), when an individual from Species 2 invades, which is capable of engaging in dormancy. Since its reproductive rate is larger than its overall death rate (including natural mortality and competition-induced mortality), the abundance of Species 2 increases, reaching a small but notable size where it begins to contribute to overcrowding, and from where any initial stochastic fluctuations are minimal ($\varepsilon$). In Phase II, competition becomes more intense due to overcrowding, which results in the decreased abundance of Species 1. However, since individuals belonging to Species 2 can temporarily escape into dormancy, they are less sensitive to competition. This allows Species 2 to continue growing towards its equilibrium population size $\bar{n}_{2a} = (\lambda_2 - \mu)/(1 - p)$, which is larger than $\bar{n}_1$ since they can tolerate higher levels of competition. In Phase III, Species 2 is close to its equilibrium. Meanwhile, since $\bar{n}_{2a} > \bar{n}_1$, Species 1 is driven towards critically low abundances ($\varepsilon$) where stochasticity again becomes visible, leading to ultimate extinction. More detail can be found elsewhere[103].

Species interactions with a seed bank: Seed banks modify species interactions with consequences for community stability and the maintenance of biodiversity. Seed banks contribute to coexistence by preventing the competitive exclusion of species through a process known as the "storage effect"[105]. First, this theory requires that species respond differently to fluctuations in environmental conditions, giving rise to what is known as temporal niche partitioning. For example, some desert plant species are adapted to wet conditions in the early winter season, while other species are adapted to drier conditions typical of early summer[106]. The second criterion stipulates that competition must covary with environmental conditions. In this case, the dry-adapted species should not be affected by competition with the wet-adapted species in the winter season. Instead, the wet-adapted species should experience stronger self-limitation, while the dry-adapted species enjoys a growth advantage when rare, which is referred to as a negative frequency dependence, a hallmark property of stabilizing coexistence. Third, the storage effect requires buffered growth, which means that, under optimal conditions, a species invests in long-lived individuals (e.g., dormant seeds) that help a population persist through periods of unfavorable conditions. The importance of the storage effect as a stabilizing mechanism of biodiversity has been well documented in a range of systems including plants[107], zooplankton[108], and protists[109].

By serving as a refuge, dormancy can modify species interactions involving predators and parasites[110,111] with implications for community stability. For example, in classical predator-prey models, an increase in the carrying capacity can promote oscillations that lead to chaotic dynamics and extinction. Such outcomes, commonly known as the paradox of enrichment, can be averted in models when predators are allowed to engage in dormancy[112], a finding that is consistent with observations of food webs involving phytoplankton prey and *Daphnia* grazers that are capable of producing ephippia[113] (Fig. 1). Some theory even suggests that internal fluctuations caused by antagonistic interactions can promote the evolution of seed banks[114]. Other lines of evidence from microbial systems tends to support the notion that seed banks modify antagonistic interactions. For example, just the physical contact between a virus and its archaeal host can initiate transitions into dormancy[115]. Meanwhile, some species fall into dormancy when they detect the chemical cues that are released by predators in their environment[116,117], a mechanism that could promote anticipatory transitioning behavior (Box 1).

Seed banks may also play a role in symbioses, including mutualisms where there is beneficial exchange of resources or services among species. Often thought to be easily disrupted, mutualistic dynamics depend on the degree to which individuals cooperate vs. cheat. Theory suggests that dormancy may stabilize mutualistic interactions, but this effect is sensitive to the time scale on which organisms transition between metabolic states[118]. If a potential partner engages in dormancy, then there should be more opportunities to initiate, establish, and sustain a long-term mutualistic relationship with another species. Yet, there are few studies that rigorously document the importance of dormancy for the maintenance of mutualisms. Lichens are a composite organism made up of mutualistic phototrophs (cyanobacteria or algae) and fungi that persist through extreme conditions (desiccation, starvation, freeze-thaw) by entering a dormant state[119,120]. Some classic plant-microbe mutualisms may also depend on dormancy. Certain strains of nitrogen-fixing rhizobia will only associate with plant varieties that engage in dormancy[121], while the germination of some plant seeds requires a fungal partner[122]. Recently, it has been shown that microbiomes can play a critical role in provisioning hosts during periods of dormancy with consequences for development. For example, gnotobiotic wasps that lack gut bacteria have reduced body mass, glucose levels, and glycerol concentrations during diapause compared to insects containing a diverse microbiome[123]. When taken together, a more comprehensive view of species interactions suggests that there may be overlooked mechanisms that offset the costs of dormancy, which might explain the prevalence of seed banks in nature.

**Box 4. Seed banks and biodiversity**

One of the most ubiquitous aspects of biodiversity is that there are very few common species and many rare species. This pattern, which is found among plants, invertebrates, mammals, and microorganisms, can be formally represented as the species abundance distribution (SAD). The SAD has attracted attention from ecologists for decades, and has inspired debate about the most appropriate statistical model to explain the classical hollow-shaped curve[124]. Variation in the SAD has led to speculation about underlying mechanisms, including the partitioning of niches and the role of disturbance events[124]. However, theory has shown that a typical SAD can also emerge emerge under neutral expectations, when there are no differences among species[125]. Understanding what gives rise to this relationship is important because many other phenomena can be derived from the SAD, including the scaling law that describes how diversity changes with increasing habitat size, a pattern known as the species area relationship (SAR).

Because seed banks affect population sizes and persistence of species over time, it is possible to explore how dormancy influences the SAD using neutral theory[125]. In the absence of dispersal limitation, the law of a sample is given by Ewen's Sampling Formula (ESF)[126]. Besides the sample size ($S_n$), the single parameter in this formula is the biodiversity constant $\theta$, which can be linked to the underlying speciation rate[127]. From the ESF, species richness of the sample ($S_n$) is sufficient for estimating $\theta$, and its expected value is given by $E[S_n] = \sum_{k=1}^{n}[\theta/(\theta + k - 1)] \sim \theta \log n$.

A particularly valuable feature of the ESF is that it can be derived from coalescent theory, and that samples can easily be generated using the Kingman coalescent[128]. Note that, in this interpretation, the universal biodiversity constant ($\theta$) agrees with the coalescent mutation rate ($\theta$) as defined in Box 2. It is thus natural to consider the effect of seed banks on the SAD and the universal biodiversity constant through their underlying coalescent structure. For example, the weak seed bank model[88] can be used, where the underlying coalescent is stretched by a factor $\beta^2$ when the expected number of generations that individuals spend in the seed bank is given by $\beta$. Thus, if a hypothetical species without a seed bank can be described by a biodiversity constant $\theta$, then a similar species with a seed bank can be described in terms of the modified constant given by $\theta^{SB} = \theta\beta^2$. This translates into a new expression for the expected species richness in the presence of a weak seed bank given by $E[S_n^{SB}] \sim \theta^{SB} \log n$.

From this, it follows that already weak seed banks can significantly alter the shape of the SAD, increasing both species richness ($S$) and the number of rare species in a way that can be explicitly computed from the ESF and the above formulas. Even when individuals only stay dormant on average for two generations, the expected richness with a seed bank ($E[S_n^{SB}] = \theta^{SB}$) increases by a factor of four. Such coalescent-based reasoning can help to explain why separate models are needed to fit the SAD for different groups of organisms. For example, the log-series distribution implemented through the maximum entropy theory of ecology (METE) does an excellent job of describing the SAD for plants and animals[129], while the log-normal distribution is better for describing diversity among microbial taxa[130]. This discrepancy has been attributed to the fact that microbial samples typically contain many more individuals ($N$) than plant and animal communities, which can lead to uneven SADs with longer tails of rare species. An alternate explanation is that seed banks influence the SAD. If so, then this effect should be pronounced in microbial communities, because seed banks are prevalent in these systems [12] and active and dormant cells are hard to differentiate. Although the ESF is not valid for predicting the effects of strong seed banks (Box 2), explicit recursive sampling formulas can be derived[78] that may yield even more pronounced effects on the SAD.

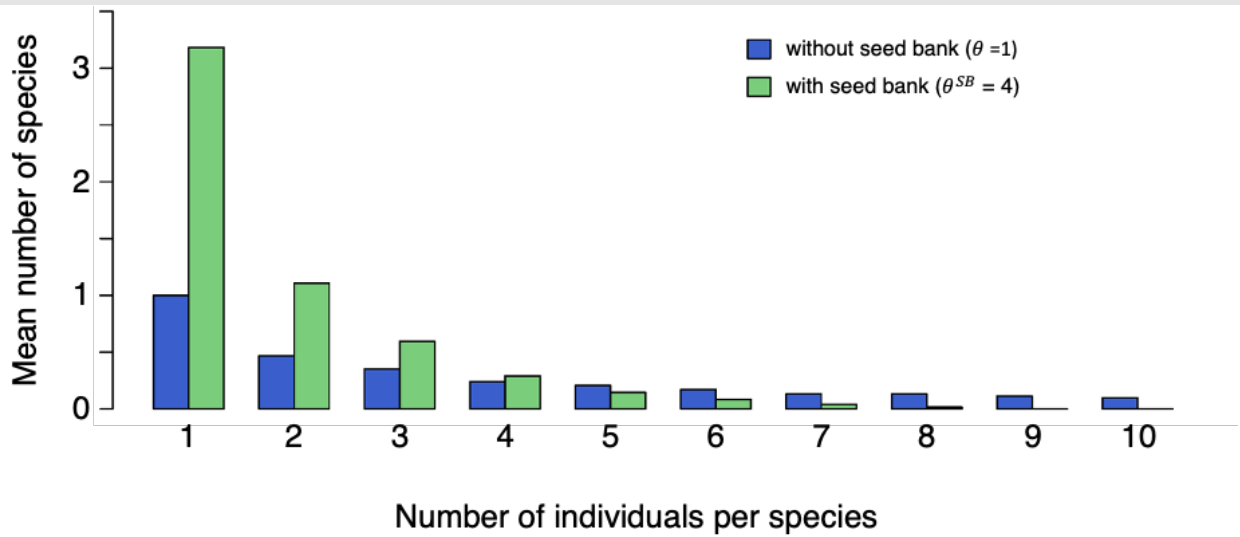

Seed banks affect the species abundance distribution (SAD) derived from the neutral theory of biodiversity[125] coupled with coalescent approaches. With a weak seed bank ($\theta^{SB} = 4$), where individuals spend on average two generations in a dormant state, species richness is higher and there is more rarity, compared to communities without a seed bank ($\theta = 1$). The figure represents SADs with a sample size of $n = 10$ averaged over 1000 iterations.

Biogeography with a seed bank: Biogeography seeks to understand how contemporary processes and historical events determine the spatial distribution and abundance of species. Thus, a major focus of biogeography is dispersal, a strategy that evolves in spatially variable landscapes. In contrast, dormancy has been overlooked in most biogeographical frameworks even though the two strategies are not independent. Despite having different risks and benefits, dormancy and dispersal are substitutable under some conditions. For example, modeling efforts have demonstrated that in environments with positive temporal correlation dormancy can increase the optimal dispersal distance of a population[131]. Such observations have inspired investigations exploring the joint evolution of dormancy and dispersal[132].

While there is evidence to support the commonly held assumption that there is a trade-off between dispersal and dormancy, this is not a universal pattern. For example, neutral and positive correlations between dispersal- and dormancy-related traits have been reported. Such relationships may arise when dormancy facilitates dispersal[65]. Not only do anatomical features of dormant life stages assist with the passive movement of individuals in a patchy landscape, they may also increase the odds of surviving harsh conditions during transit. Finally, upon arrival in a new environment, dormancy may provide an extended window of opportunity for conditions to arrive that will trigger resuscitation, which in turn should increase the probability of successful colonization[65].

Growing evidence suggests that biogeographic patterns are influenced by seed-bank dynamics. When explicitly incorporated into mathematical models, local and regional patterns of diversity are dependent on the degree to which dormancy and dispersal covary[65]. Empirical studies further support the view that dormancy is an important process that should be incorporated into biogeographical frameworks. For example, in a study of bacterial communities sampled from a network of forested ponds, molecular-based procedures were used to distinguish sequences that belonged to active and inactive individuals[67]. In combination with process-based models, researchers were able to evaluate the relative importance of dormancy and dispersal to a common biogeographic pattern known as the distance-decay relationship (DDR). The DDR describes how the compositional similarity of samples made up of different species (Fig. 2) decreases with increasing geographic distance. It has been shown that slopes describing the DDR for microorganisms tend to be quite shallow compared to those of plant and animal communities[133]. A reasonable explanation for this distinction is that microorganisms have more cosmopolitan distributions owing to their small size, which allows them to be more easily transported over long distances. By explicitly accounting for variation in metabolic activity among microbial taxa, an alternate explanation is that spatial patterns of biodiversity can be affected by dormancy and seed-bank dynamics[134].

**Box 5. Spatial seed banks**

In addition to affecting local population-genetic processes (Box 2), seed banks can affect patterns of diversity at larger spatial scales. Consider a metapopulation where each individual belongs to a colony (or patch) that has a fixed population size within a landscape of colonies. Active individuals can migrate between colonies and undergo clonal reproduction via resampling within a colony. Each of the colonies contains a structured seed bank. Upon entering the seed bank, dormant individuals forgo the opportunity to reproduce or migrate. Instead, those individuals are randomly assigned to a compartment ($D_0, D_1, \ldots, D_m, \ldots$), which specifies the amount of time they will spend in the seed bank prior to resuscitation. While preserving the Markov property of the evolution, this structuring of dormant individuals allows one to explore how the distribution of wake-up times affects genetic diversity in a spatially explicit manner. The system can be modeled via a set of coupled stochastic differential equations, describing the population in the large-colony-size limit.

The type of equilibrium the system is driven into depends on the interplay between dormancy and migration[135]. In the absence of a seed bank, the colonies converge to a joint equilibrium whose nature is solely determined by migration. If $p_t$ denotes the probability that at time $t$ a lineage has returned to the colony it originated from, then a monotypic equilibrium emerges if and only if the migration is recurrent: $\int_0^\infty p_t dt = \infty$. This means that, sooner or later, each lineage will return to its original colony. Looking back in time, if two or more lineages repeatedly meet in the same colony, then there is a considerable chance that they will have the same ancestor and thus coalesce (Box 1). As a result, their descendants will possess the same genotype in the contemporary population.

In the presence of a seed bank, the system is more likely to converge on a polytypic equilibrium. This is because dormancy creates a much more restrictive set of conditions for a monotypic equilibrium: $\int_0^\infty t^{-(1-\gamma)/\gamma} p_t dt = \infty$. Here, $\gamma$ controls the tail of the wake-up time from the seed bank in the sense that the probability for the resuscitation time to exceed $t$ falls off like $t^{-\gamma}$ for large $t$. When $\gamma < 1$, this implies that the wake-up time has infinite mean. For the integral to be infinite, migration not only needs to be recurrent, but lineages actually need to spend drastically more time at their original colony. Again, looking back in time, two lineages in a colony with a seed bank can coalesce and thus share the same genotype in the present. But this will only occur if both lineages are active at the same time, which becomes increasingly unlikely with decreasing probabilities of resuscitation (i.e., small values of $\gamma$). Taken together, migration and seed-bank structure have important implications for the genetic diversity of populations at the landscape scale.

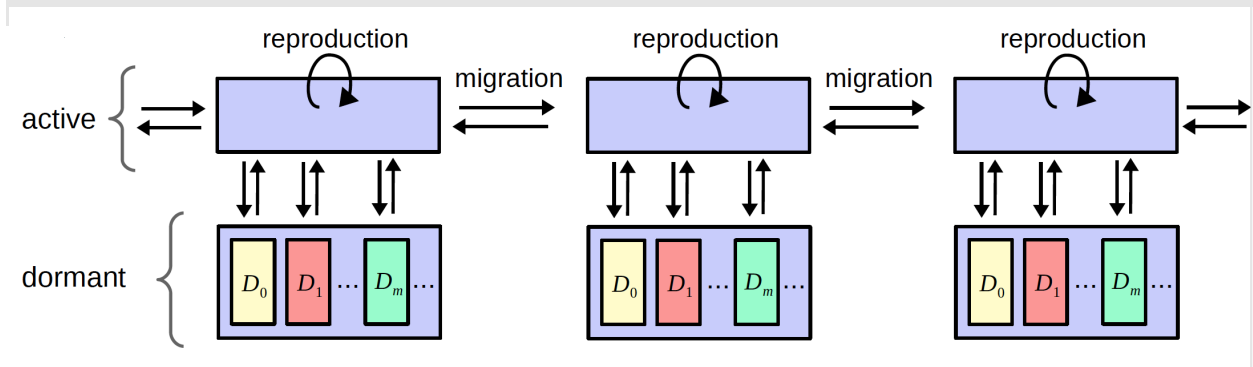

Spatial model to understand genetic diversity with a seed bank. Individuals are subject to migration (dispersal) and resampling (reproduction). They can also move into and out of the seed bank (initiation of dormancy and resuscitation). When individuals enter a dormant state ($D$), they are randomly assigned a compartment ($m$ = 0; 1; …), which determines their time until resuscitation.

## FRONTIERS

After outlining the fundamental attributes and transitions that control seed banks, and describing how dormancy dynamics can give rise to emergent phenomena, we now switch gears to discuss opportunities, applications, and frontiers for seed-bank theory. It is not our goal to comprehensively review the details of the major themes that follow. Rather, our aim is to identify instances where seed-bank theory can be used to better inform and quantitatively address complex problems in the life sciences. Some of these advances may be readily achieved using existing technology and resources, while others will require the development of new theories, modeling, and statistical approaches resulting from collaborations among mathematicians, physicists, computer scientists, and biologists.

**Human health:** Most theoretical and empirical efforts related to seed banks have focused on the evolution and ecology of environmental systems. However, the principles of seed banks are directly relevant to the persistence, pathogenicity, and spread of diseases, and therefore may be fruitful for developing novel therapeutic treatments to improve human health.

<u>Chronic infections</u> — Many infectious diseases are caused by microorganisms that have the capacity to persist for extended periods of time inside of hosts via dormancy[136]. While in a metabolically inactive state, pathogens can escape host immune defenses, but they are also less susceptible to commonly used drugs that target the transcriptional and translational machinery of growing cells. As a consequence, microbial dormancy may contribute to the rise and spread of antibiotic resistance[137]. For example, there are many strains of clostridia that inhabit the mammalian gut, where they provide benefits to their hosts by attenuating inflammation and producing metabolites like short-chain fatty acids that help regulate neuro-immuno-endocrine networks. However, some strains, like *Clostridium difficile*, are gastrointestinal pathogens that are difficult to eliminate using traditional drug therapies. In part, this is because *C. difficile* forms endospores, which not only provide protection against antibiotics, but also facilitate transmission among hosts[138]. Other pathogens lacking the ability to form spores can survive drug exposure through their association with biofilms[139]. These multicellular structures reduce the diffusion of drugs, but biofilms can also serve as a reservoir of persister cells. Recognized as a subpopulation of non-growing cells that are tolerant to antibiotics, many persisters stochastically enter dormancy via mechanisms involving toxin-antitoxin modules, while microfluidics and time-lapse microscopy have demonstrated that low-energy (ATP) cells are capable of surviving exposure to ampicillin[140]. Such knowledge may be useful for developing novel therapeutics to treat chronic infections, including *Mycobacterium tuberculosis*, the causative agent of TB. Many variants of *M. tuberculosis* are difficult to monitor and treat because dormant cells escape diagnostic tests that are commonly used by clinicians. However, these so-called "occult" populations can be awoken by using resuscitation promoting factors (Rpf), which are muralytic proteins that terminate dormancy by cleaving bonds of peptidoglycan found in the cell wall [141]. Once dormant pathogens are activated, they should be easier to treat and eliminate using conventional drug therapies. Together, these examples demonstrate how chronic infections can be better understood and managed using the principles of seed banks.

Cancer — A non-communicable disease responsible for nearly 10 million deaths per year, cancer arises from uncontrolled growth of abnormal cells. Scientists have long approached cancer by focusing on the development, regulation, and division of somatic cells that acquire mutations and proliferate over the lifetime of humans and other species. Typically, there is an emphasis on molecular processes, including the activation and inactivation of genes that control the cell cycle and the process of apoptosis, which are collectively viewed as targets for pharmaceutical development and therapeutic treatment. In addition, recent advances in the high-throughput sequencing of genomes and epigenomes, in combination with bioinformatics and artificial intelligence, hold promise for breakthroughs in various areas of oncology and precision medicine[142]. However, the principles of ecology and evolution are also being used to understand the complexity and dynamics of cancer. For example, cancer cells have different phenotypes that reflect variation in resource consumption, stress tolerance, longevity, and growth that set the stage for trade-offs with implications for tumor persistence[143]. One feature that is universally recognized by researchers, ranging from physicians to theoreticians, is that cancer dynamics are affected by a cell's ability to engage in dormancy. Patients can be in remission for decades until cancer cells are resuscitated from a dormant state. As with plants, cancer dormancy can be achieved via different metabolic phenomena, including the non-proliferation of individuals that enter cell-cycle arrest, or quiescence where individuals exhibit slow-growing, stem-like properties[23]. Patient relapses often arise when cancerous cells transition between phenotypic states, sometimes due to noise in gene expression, a process that has been incorporated into stochastic Markov models[144]. Alternatively, cancer dormancy can involve responsive transitioning (Box 1), where cells enter and exit dormancy owing to fluctuations in environmental conditions within the host, such as cycles of hypoxia, fluctuations in levels of toxic metabolites, or variation in blood flow[145]. Evidence suggests that lineages respond to variability in these conditions in different ways, which means that ecological theory, like the storage effect, may be useful for predicting cancer dynamics[145]. Dormancy also plays an important role in metastasis, a process where cells break off from tumors at their site of origin, enter the blood stream or lymphatic system, and disperse to other parts of the body. Dormancy defends the cells from drugs and the host immune system while in transition, and on the way may facilitate the establishment of new tumors, which often express different sets of symptoms and pathologies[146]. Taken together, new and potentially effective therapeutics may be developed that take into account the principles and expectations from seed-bank theory.

Epidemiology — Because dormancy can affect spatial and temporal processes in complex ways, seed banks may be useful for predicting the spread of diseases. For example, in some long-lived hosts, asymptomatic individuals harbor covert or "silent" infections[147]. Viewed in the context of dormancy, these infections can have stabilizing or destabilizing effects on host population dynamics, especially when they switch to an overt infection, which increases horizontal transmission and the probability of outbreaks[148]. In some cases, the dynamics of covert infections may be contingent upon interactions with other species, as evidenced by the resuscitation of dormant *E. coli* in the urinary tract after exposure to members of the vaginal microbiome[149]. There are also opportunities to incorporate dormancy into classical SIR models, which track how individuals move between susceptible ($S$), infectious ($I$), and recovered ($R$) states. Although several related forms of latency and carriers states have been explored[150], a systematic analysis of dormancy in this epidemiological context is lacking. For example, SIR-like models could incorporate dormancy-related processes into host population, pathogen populations, or both. While such models could be used in public health, they may also be valuable for understanding the dynamics of foodborne pathogens like *Listeria* that are infected by viruses (i.e., phage). For example, CRISPR-based immunity can inactive RNA-based viruses, but it also shuts down the metabolic activity of hosts that harbor the pathogen-derived spacers. Meanwhile, other bacteria in the population lacking the appropriate CRISPR spacer will indirectly benefit from the clearance of the virus. In this way, dormancy may help confer herd immunity to the bacterial population with implications for controlling some epidemics[151].

**Global change**: Environmental change is occurring on a global scale at an unprecedented rate. Increasing temperatures, altered precipitation, increased nutrient runoff, and habitat destruction pose major threats to the long-term persistence of species, while also altering the biogeochemical cycles with implications for the functioning of the biosphere. In addition to helping some species contend with current and future global change scenarios, seed banks may be useful for forecasting ecosystem dynamics.

<u>Conservation and restoration</u> — Many natural and managed ecosystems benefit from seed banks. Reservoirs of dormant individuals can stabilize populations and contribute to the predictable pattern of community succession following disturbance events. A classic example is the recovery of terrestrial ecosystems following fire events, which can be a significant source of mortality for adult plants. Many species contend with these recurrent disturbances through the evolution of serotiny, a common trait where the release and germination of seeds requires the heat of a fire[152]. However, seed banks are not a universal safeguard for maintaining species in the face of global change. For example, in the Arctic, where environmental conditions are accelerating most rapidly, seed-banking is not a universal safeguard for all plants species. In a survey of native angiosperms, seeds from 40 % of species failed to germinate; these observations were disproportionately associated with rare taxa, which have a greater risk of extinction[153]. Similarly, in arid regions of Australia, seeds from three out of eight plants species had higher rates of germination after exposure to elevated temperatures, but other species were less resilient to the simulated global-change scenario[154]. Although seed banks are often viewed as beneficial, they can also interfere with some conservation and restoration goals. For example, invasive species can create a positive feedback by enriching the local seed bank with propagules that promote the spread of taxa throughout the landscape[155]. Under such conditions, the removal of adult plants, which is a common management strategy, may have minimal to no effect on controlling invasive species owing to recruitment from the belowground seed bank. Similarly, recent studies showed that waking up microorganisms from the seed bank can have negative effects on co-occurring plants, perhaps owing to the dilution of mutualists or the awakening of pathogens[156]. Meanwhile, concerns were raised about the re-emergence of pathogens from the past when the thawing of Siberian permafrost led to a die-off of reindeer that were infected with a putatively ancient strain of *Bacillus anthracis*[157].

Ecosystem functioning — Seed banks have implications for the functioning and stability of entire ecosystems. Owing to their abundance and diversity, microorganisms play a particularly important role when it comes to the flux of materials and energy in food webs. Yet, a non-trivial fraction of the microbial community is made up of metabolically inactive microorganisms[12], which raises questions about how seed banks contribute to global change. Decades of nutrient enrichment altered the size and composition of microbial seed banks in an estuarine ecosystem[158]. In this way, a disruption of nutrient cycles at the land-water interface could contribute to the build-up of seed banks and the emergence of harmful algal blooms, which result in fish die-offs, diminished drinking-water quality, and altered biogeochemistry. Such thinking has inspired scientists to incorporate traits like dormancy into statistical and process-based models, with the hope of better understanding the complex ways in which microbes influence Earth system processes. For example, in a global change study where drought and temperature were experimentally manipulated in a grassland ecosystem, soil respiration could be better explained when the fraction of dormant bacteria was explicitly taken into account[159]. When applied to Arctic ecosystems, models predict that substantially more carbon will be stored in soils owing to microbial dormancy, which has important implications for understanding microbial processes at local, regional, and global scales[160].

**Box 6. Mathematical frontiers**

The mathematical treatment of seed banks has been expanding over the past decade. Originally motivated by biological observations, the objects and properties of seed banks are now opening up important questions and approaches in various subfields of mathematics, including complex interacting systems. In this context, dormancy can be seen as a new paradigm with the capacity to alter the behavior of classical models across multiple scales by introducing memory and delays through the introduction of feedback. In turn, the principles and tools of mathematics can fertilize research in the social and life sciences. Here, we briefly outline questions and opportunities that will benefit from the mathematical analysis of seed banks.

Interacting particle systems — At the interface of probability theory and statistical physics, the field of interacting particle systems focuses on collective behaviors and phase transitions in systems consisting of a large number of interacting components. Key techniques include graphical representation, coupling, duality, and coalescents[161,162]. The theory allows particles to exist in various states (e.g., present/absent, up/down), but also permits them to move through space. Thus, a natural extension would be to add dormancy to interacting particle systems. For example, a particle could fall asleep, remain idle for a period of time, and then resume spatial motion and interaction with other particles after waking up. Particles may also undergo metabolic transitions, which would make them behave differently than particles that consistently remain active. In such ways, dormancy may affect the collective behavior of particles and give rise to new types of phase transitions. Multi-type interacting particle systems represent a rich modelling class and typically have complicated duals, capable of incorporating a multitude of phenomena. The interaction between different species of particles may result in a violation of transport laws that are classical for a single species of particles, including Fourier's law for currents across density gradients. For instance, interaction between fast and slow particles may result in a current that goes against the gradient, which is remarkable.

Complex networks — Many natural and engineered systems are organized into networks. Often large and complex, they represent connectedness and spatial relationships between different objects in social, technological, economic, and biological systems. As a result, network-based approaches have been used in a range of quantitative disciplines, including graph theory, combinatorics, probability theory, and computer science[163]. Researchers have begun to inquire about aspects of seed banks by investigating the importance of active and dormant connections in social networks[164,165]. The human brain is another example of a network that has seed-bank properties. For example, it appears that dormant neurons spike more slowly than active neurons[166].

Such observations are consistent with the notion that external events can cause dormant psychopathologies to become activated and trigger other mental disorders[167]. Despite advances in these areas, it remains unclear how the form and functionality of networks are influenced by the presence of dormancy and seed banks. In particular, spatially heterogenous seed banks, modulated by an underlying network structure, represent one such challenge for understanding network function. It is possible that there are common motifs (e.g., autoregulation, feed-forward, and feed-back) involving seed banks that can be leveraged for robustness and adaptability in networks. One goal is to determine whether new universality classes of collective behavior occur in a network-based framework that are different from those found in other seed-bank structures.

Adaptive dynamics — Adaptive dynamics is a theoretical approach that describes changes in trait distributions in heterogeneous populations[168]. It considers aspects of population genetics (e.g., hereditary mechanism, mutation) and population dynamics (e.g., competition, fitness), while assuming a separation between ecological and evolutionary time scales. Originally a deterministic platform, adaptive dynamics now includes stochastic individual-based models, which together exhibit a rich scaling behavior that can be described by measure-valued diffusions[169]. Introducing memory in the form of a seed bank should affect many aspects of these scaling limits in important ways, including the dynamics of the trait substitution sequence (TSS), the polymorphic evolution sequence (PES), or the canonical equation of adaptive dynamics (CEAD)[170-172]. In regimes with strong selection and weak mutation, so-called adaptive walks appear as scaling limits, moving in fitness landscapes[173]. Again, dormancy and the resulting reservoirs of genotypic and phenotypic variability may affect the behavior of these systems, in particular if fitness landscapes change over time. Many additional mechanisms interacting with dormancy are waiting to be studied, including horizontal gene transfer, predation, or cancer dynamics.

Stochastic partial differential equations — Seed banks can be incorporated into classical stochastic partial differential equations. This can be achieved by adding a dormant component where at least one of the driving forces is deactivated, thus allowing for continuous two-way migration between the two components. Such an explicit expansion would allow for delay-representation, leading to an explicit description of the memory introduced by a seed bank. Concrete effects of a seed bank have already been documented by using the stochastic Fisher-KPP equation, where the addition of a dormant state significantly reduces the speed at which beneficial alleles spread through the population[174]. Other seed-bank mechanisms can be expected to manifest themselves in stochastic partial equations if they can be derived from microscopic systems that allow for dormant states. For example, one may speculate about the effects of dormancy on the KPZ-equation arising as a scaling limit in models for bacterial growth or the exclusion process.

## CONCLUSION

We make the case that seed banks play a crucial role in determining the behavior observed in a variety of biological systems, spanning collections of cells to entire ecosystems. Seed banks come with different attributes and physical features, each of which can affect the transitions into and out of dormancy in complex ways, giving rise to a multitude of emergent phenomena. The mathematical description of seed banks is in many areas still in its infancy, but various new techniques are being developed and new challenges are being identified. We argue in favor of a systems theory perspective, in which different lines of research and different viewpoints are combined to facilitate further progress. Seed banks represent a paradigm that will keep us busy for decades to come.

## ACKNOWLEDGEMENTS

We acknowledge support from the National Science Foundation (DEB-1442246 and 1934554 JTL), US Army Research Office Grant (W911NF-14-1-0411 JTL), the National Aeronautics and Space Administration (80NSSC20K0618 JTL), the Netherlands Organisation for Scientific Research through NWO Gravitation Grant (NETWORKS-024.001.003 FdH), the Alexander von Humboldt Foundation (FdH), the German Research Foundation (DFG SPP 1590, Probabilistic Structures in Evolution), and the Berlin Mathematics Research Center MATH+ (JB and MWB). The manuscript was improved with feedback from F. Baumdicker, A. Magalie, D. Schwartz, P. Wall, N. Wisnoski, and from four anonymous reviewers and the editor.